\documentclass[nofootinbib,reprint,amsmath,amssymb,aps]{revtex4-1}

\usepackage{graphicx}
\usepackage{dcolumn}
\usepackage{bm}
\usepackage{hyperref}
\usepackage[mathlines]{lineno}

\begin{document}

\preprint{APS/123-QED}

\title{Bianchi identities in $f(T)$ gravity:\\
paving the way to confrontation with astrophysics}

\author{Alexey Golovnev}
\altaffiliation[]{agolovnev@yandex.ru}
\affiliation{Centre for Theoretical Physics, The British University in Egypt, 11837 El Sherouk City, Egypt}

\author{Mar\'ia-Jos\'e Guzm\'an}
\altaffiliation[]{maria.j.guzman.m@gmail.com}
\affiliation{Departamento de F\'isica y Astronom\'ia, Facultad de Ciencias, Universidad de La Serena, Av. Juan Cisternas 1200, 1720236 La Serena, Chile}

\date{\today}

\begin{abstract}
Theories of $f(T)$ gravity are being actively confronted with cosmological observations, and are being studied for their potential to solve famous problems of cosmology. A necessary step is to extend these studies to astrophysical settings. However, to this end one must understand the structure of spherically symmetric solutions. We show that two different known approaches to these solutions are actually fully equivalent from the point of view of Lorentz-covariant formalism. Moreover, we explain Bianchi identities in $f(T)$ gravity and apply them to show that the corresponding equations are always compatible. It puts these efforts on much firmer grounds than before.

\end{abstract}

\maketitle

\section{\label{sec:intro} Introduction}

One of the recurrent themes in modern theoretical physics is the crisis in the $\Lambda$CDM standard model of cosmology  \cite{Bull:2015stt}. Not only do we not know the nature of the Dark Sectors, or do we face problems at highly non-linear astrophysical scales \cite{DelPopolo:2016emo}, but also direct inconsistencies appear when trying to explain current data, such as the recent tension between different measurements of $H_0$  \cite{Bernal:2016gxb}. They are related to regimes in which we are confident that our understanding of physics should work well.

This crisis motivates numerous attempts at alleviating the problems by a proper modification of Einstein's general theory of relativity. The history of these modifications, though dating back to Einstein himself, does not flourish with success, and together with other problems of general relativity such as the lack of commonly accepted quantum gravity theory despite many decades of hard work, it leads to necessity of revising the geometrical foundations of gravity \cite{BeltranJimenez:2019tjy}.

One of the oldest ideas in this direction is the teleparallel approach to gravity which brought the $f(\mathbb T)$ gravity models to spotlight some years ago \cite{Ferraro:2006jd,Bengochea:2008gz,Ferraro:2008ey}. They can naturally provide phantom regimes and allow to build  background cosmologies \cite{Awad:2017sau} capable of resolving the $H_0$ tension \cite{Nunes:2018xbm,El-Zant:2018bsc,Wang:2020zfv}, which explains why they have become very popular in cosmological model building \cite{Cai:2015emx}.

Due to their importance in the field of theoretical cosmology, modified teleparallel theories of gravity attract much interest as prospective alternatives for description of gravitational interactions. Of course, it is necessary to confront them with observations also beyond the realm of cosmology. In particular, astrophysics provides natural motivation for studying spherically symmetric solutions, which are a tool that can hardly be overestimated when it comes to deriving observable effects of gravity.

Up to now, the status of spherically symmetric solutions in $f(\mathbb T)$ gravity has been rather unclear. Vacuum (Black Hole) solutions were discussed already some time ago \cite{Ferraro:2011ks,Nashed:uja,Nashed:2013bfa,Nashed:2014sea,Bejarano:2014bca,Paliathanasis:2014iva,Bejarano:2017akj}, however those were solutions with constant (zero) torsion scalar. At the background level such solutions are the same as in general relativity with a cosmological constant. This is an ad hoc construction to make the torsion scalar constant in order to reduce the equations to the simple general relativistic case. The corresponding tetrads usually look very contrived and have very little respect to the symmetry of the metric.

More interesting are solutions with non-constant torsion scalar. Only a few of them are known exactly for special choices of the function $f$, found by heavy mathematical machinery \cite{Paliathanasis:2014iva,Bahamonde:2019jkf,Boehmer:2019uxv,Boehmer:2020hkn}. However, generically these solutions are derived in terms of series expansion \cite{Bahamonde:2019zea,Bahamonde:2020bbc}. An issue with them is that the problem is presented as a system of three different equations, with no obvious reason for dependence among each other, for only two functions. As far as we know, no explanation of why this system is solvable was offered in existing literature. One can suspect that the known exact solutions are very special, and perturbation theory might not capture all potential problems in other cases. In private communications it was even possible to hear an opinion that $f(\mathbb T)$ gravity lacks spherically symmetric solutions in general.

The main aim of this Letter is to show that the missing link in the chain are the Bianchi identities. They exist in $f(\mathbb T)$ gravity, though in a modified form, and they allow us to show that the three equations are indeed compatible. It means that we have proven that spherically symmetric solutions do generically exist, henceforth establishing firm grounds for testing modified teleparallel gravity models against astrophysical observations. 

The outline of this Letter is as follows. We provide a concise summary on teleparallel geometry in Sec. \ref{sec:mtg}, alongside introducing $f(\mathbb T)$ gravity. Our main result, that Bianchi identities exist in $f(\mathbb T)$ gravity, is presented in Sec. \ref{sec:bianchi}, where we exhibit the proof of the identity in a formalism-independent way. We recall the classical spherical symmetric tetrad ans\"atze in Sec. \ref{sec:spherical}, for both tetrad-based and covariant formalisms and their equations of motion, and argue that the two tetrads are merely related by a Lorentz rotation. In Sec. \ref{sec:exist} we present how the Bianchi identities allow to prove the existence of spherically symmetric solutions in $f(\mathbb T)$ gravity. We also discuss the physical feasibility of very simple exact solutions. We finally settle our conclusions in Sec. \ref{sec:concl}.

\section{Modified teleparallel gravity}

\label{sec:mtg}

The teleparallel description of gravity is based on a flat metric-compatible spacetime connection \footnote{The decomposition of the full connection including the contribution from non-metricity can be found in Refs.\cite{Ortin2015,Jimenez:2020dpn}}
\begin{equation}
\label{conn}
\Gamma^{\alpha}_{\mu\nu}=e_a^{\alpha}\left(\partial_{\mu}E^a_{\nu}+\omega^a_{\hphantom{a}\mu b}E^b_{\nu}\right)    
\end{equation}
where $e^{\mu}_a$ are the components of the tetrad field, $E^{a}_{\mu}$ its inverse, and $\omega^a_{\hphantom{a}\mu b}$ is a flat spin connection,
\begin{equation}
\label{inert}
\omega^a_{\hphantom{a}\mu b}=-(\Lambda^{-1})^c_b \partial_{\mu}\Lambda^a_c
\end{equation}
with an arbitrary Lorentz matrix $\Lambda (x)$ \cite{Golovnev:2018red}. 

The original formulation of teleparallel geometry considers the Weitzenb{\"o}ck choice of $\omega=0$ because of its simplicity, and this is indeed a possible way to fix the freedom allowed by the equations of motion \cite{Bejarano:2019fii} of covariant version. However the so-called inertial spin connection \eqref{inert} can equally well describe a flat and metric-compatible spacetime, and it plays a significant role later in nonlinear modifications of teleparallel gravity \cite{Krssak:2015oua,Golovnev:2017dox,Hohmann:2018rwf,Ferraro:2018tpu}.

The torsion tensor is the antisymmetric part of the connection \eqref{conn}, that is $T^{\alpha}_{\hphantom{\alpha}\mu\nu}=\Gamma^{\alpha}_{\mu\nu}-\Gamma^{\alpha}_{\nu\mu}$, and it is the main object with which we define the action of the teleparallel equivalent of general relativity (TEGR):
\begin{equation}
\label{tegr}
    S = - \dfrac{1}{2\kappa} \int d^4 x E\cdot \mathbb T \equiv  - \dfrac{1}{2\kappa} \int d^4 x E\cdot T_{\alpha\mu\nu}S^{\alpha\mu\nu},
\end{equation}
where the torsion scalar $\mathbb T$ is implicitly defined by the second equality, the integration measure $E=\text{det}(E^a_{\mu})$ is the determinant of the inverse tetrad, $\kappa=8\pi G$, and 
\begin{equation}
\label{spot}
S^{\alpha\mu\nu}=\frac12\left(K_{\mu\alpha\nu}+g_{\alpha\mu}T_{\nu}-g_{\alpha\nu}T_{\mu} \right)
\end{equation}
is known as the superpotential. Here $T_{\mu}\equiv T^{\alpha}_{\hphantom{\alpha}\mu\alpha}$ is the torsion vector, and
\begin{equation}
\begin{split}
K_{\alpha\mu\nu} & = g_{\alpha\beta} \left(\Gamma^{\beta}_{\mu\nu} - \mathop{\Gamma^{\beta}_{\mu\nu}} \limits^{(0)}\right) \\
& =\frac12 \left(T_{\alpha\mu\nu}+T_{\nu\alpha\mu}+T_{\mu\alpha\nu}\right)
\end{split}
\end{equation}
is the contortion tensor. $\mathop{\Gamma^{\beta}_{\mu\nu}} \limits^{(0)}$ is the standard Levi-Civita connection expressed in terms of the metric $g_{\mu\nu}$, so all quantities that have a symbol ${(0)}$ at the top are calculated with this connection. The TEGR model is equivalent to GR because $\mathbb T$ is equal to minus the Levi-Civita scalar curvature modulo a surface term \cite{Pereira2012}.

The $f(\mathbb T)$ gravity action is the simplest nonlinear generalization of the TEGR action, given by
\begin{equation}
S = - \dfrac{1}{2\kappa} \int d^4 x E\cdot f(\mathbb T).
\end{equation}
The equations of motion of this action are obtained by deriving with respect to the tetrad field. Denoting $f_T\equiv\frac{df}{d\mathbb T}$ and $f_{TT}\equiv\frac{d^2f}{d\mathbb T^2}$, these can be conveniently written as
\begin{equation}
\begin{split}
0=\mathfrak T_{\lambda}^{\hphantom{\lambda}\nu} & \equiv \kappa \Theta^{\nu}_{\lambda} + 2f_{TT}(\mathbb T)  S_{\lambda}^{\ \mu\nu}\partial_{\mu}\mathbb T - \dfrac{1}{2} \delta^{\nu}_{\lambda} f(\mathbb T) \\
& + 2e E^{a}_{\lambda}f_T(\mathbb T) {\mathcal D}_{\mu}[E e^{\sigma}_a S_{\sigma}^{\ \mu\nu}  ] + 2T^{\rho}_{\ \mu\lambda} S_{\rho}^{\ \mu\nu} f_T(\mathbb T)
\end{split}
\end{equation}
where $\Theta^{\nu}_{\lambda}$ is the stress-energy tensor for a matter field, $\mathcal D$ is the Lorentz-covariant derivative, in particular ${\mathcal D}_{\mu}e^{\sigma}_a=\partial_{\mu} e^{\sigma}_a-\omega^b_{\hphantom{b}\mu a}e^{\sigma}_b$. In the Weitzenb{\"o}ck gauge it coincides with the usual partial derivative. Remember that the usual covariant derivative on spacetime indices is
$\mathop{\bigtriangledown_{\mu}}\limits^{(0)} A_{\nu}=\partial_{\mu}A_{\nu} - \mathop{\Gamma^{\alpha}_{\mu\nu}} \limits^{(0)} A_{\alpha}$.

A usual way of writing the equations of motion of $f(\mathbb T)$ in a covariant form is:
\begin{multline}
\label{coveom}
0=\mathfrak T_{\mu\nu}  \equiv  \kappa \Theta_{\mu\nu}-f_T({\mathbb T})\mathop{G_{\mu\nu}}\limits^{(0)}-2f_{TT}(\mathbb T)S_{\mu\nu\alpha}\partial^{\alpha}{\mathbb T}\\-\frac12 \left(\vphantom{f^A_B}f(\mathbb T)-f_T(\mathbb T){\mathbb T}\right)g_{\mu\nu}.
\end{multline}
Note that our definition of the superpotential in Eq.\eqref{spot} is different by a factor of $\frac12$ from the one adopted in some papers on the subject, hence seemingly different equations.

The interest in modified teleparallel gravity arises from the search for alternative mechanisms to explain inflation \cite{Ferraro:2006jd}, but soon $f(\mathbb T)$ gravity was used for explaining the dark energy paradigm \cite{Bengochea:2008gz}, and since then it has been extensively studied in a cosmological context \cite{Cai:2015emx}. A matter of concern in recent years comes from evidence of a strong coupling problem, since extra propagating modes \cite{Li:2011rn,Ferraro:2018tpu,Ferraro:2020tqk,Jimenez:2020ofm}\footnote{For a summary about  different approaches on the d.o.f. of $f(\mathbb T)$ gravity, see Section 3.2 of Ref. \cite{Golovnev:2019kcf}.} do not manifest around spatially flat FLRW cosmology \cite{Golovnev:2018wbh}. However, they show up at 4th order around a trivial Minkowski tetrad \cite{Jimenez:2020ofm}, which could suggest that cosmological perturbations should be studied at higher orders or backgrounds different from the trivial ones should be perturbed. It is clear that a better understanding of these issues for models beyond $f(\mathbb T)$ is needed, which can help to explore alternative modified teleparallel gravities that do not possess these pathologies and have well behaved degrees of freedom. 

\section{\label{sec:bianchi} Bianchi identities in $f(\mathbb T)$ gravity}

The issue of Bianchi identities in $f(\mathbb T)$ gravity was surprisingly ignored in previous literature, and was only briefly mentioned in the paper \cite{Golovnev:2018wbh} of one of us. Now we want to extend this important discussion a bit. 

We define equations of motion ${\mathfrak T}^{\mu\nu}$ via the variational derivative of the action functional as
\begin{equation}
\kappa\frac{\delta S}{\delta E^a_{\mu}}\equiv E {\mathfrak T}^{\mu\nu} E^b_{\nu}\eta_{ab}.
\end{equation}
Diffeomorphism invariance of the action implies that the variation
\begin{equation}
\kappa\delta S=\int d^4 x\cdot  E {\mathfrak T}^{\mu\nu} E^b_{\nu}\eta_{ab}\cdot \delta E^a_{\mu}
\end{equation}
vanishes identically if the tetrad variation is of the form
\begin{equation}
E^a_{\mu}\longrightarrow E^a_{\mu}-E^a_{\nu}\partial_{\mu}\zeta^{\nu}-\zeta^{\nu}\partial_{\nu}E^a_{\mu}.
\end{equation}
Notice that this transformation leads to
\begin{equation}
\label{Bident}
e\partial_{\mu}\left(E{\mathfrak T}^{\mu}_{\nu}\right)-{\mathfrak T}^{\lambda}_{\mu}e^{\mu}_{a}\partial_{\nu}E^a_{\lambda}=0
\end{equation}
plus a total derivative $\partial_{\mu}(E{\mathfrak T}^{\mu}_{\nu} \zeta^{\nu} )$, therefore the variation of the action vanishes up to a boundary term, therefore on-shell Bianchi identities only require the variation of the action under diffeomorphisms to be a boundary term. Eq. \eqref{Bident} can be transformed into
\begin{equation}
\mathop{\bigtriangledown_{\mu}}\limits^{(0)}{\mathfrak T}^{\mu\nu}+K^{\alpha\nu\beta}{\mathfrak T}_{\alpha\beta}=0.
\end{equation}
When the local Lorentz symmetry is not broken (we assume pure tetrad formulation here), invariance of the action under $E^a_{\mu}\longrightarrow \Lambda^a_b E^b_{\mu}$ implies that ${\mathfrak T}^{\mu\nu}$ is symmetric, and by virtue of the  antisymmetry of the contortion tensor, the usual Bianchi identities are restored. In $f(\mathbb T)$ gravity this is not the case. However, if the antisymmetric part of equations is satisfied we see that the rest must obey the Bianchi identity.

Note that one can also prove the Bianchi identities at the level of the equations of motion. Indeed, let us take divergence with respect to the index $\nu$ in covariant equations of motion (\ref{coveom}). We get, cancelling the $f_{TT}$ factor:
\begin{equation}
-\mathop{G^{\mu\nu}}\limits^{(0)}\partial_{\nu}\mathbb T-2(\partial_{\alpha}\mathbb T)\mathop{\bigtriangledown_{\nu}}\limits^{(0)}S^{\mu\nu\alpha}+\frac12 g^{\mu\nu}\mathbb T \partial_{\nu}\mathbb T
\end{equation}
where we have used the usual Bianchi identities and
\begin{equation}
\begin{split}
S^{\mu\nu\alpha}\mathop{\bigtriangledown_{\nu\alpha}}\limits^{(0)}\mathbb T & = -S^{\mu\nu\alpha}(T_{\beta\nu\alpha}-2K_{\beta\nu\alpha})\partial^{\beta}\mathbb T \\
& = S^{\mu\nu\alpha}(T_{\nu\beta\alpha}+T_{\alpha\beta\nu})\partial^{\beta}\mathbb T=0.
\end{split}
\end{equation}

To see whether the obtained relation can be proven identically zero, one can derive using $\mathop{\Gamma^{\beta}_{\mu\nu}}\limits^{(0)}=\Gamma^{\beta}_{\mu\nu}-K^{\beta}_{\hphantom{\beta}\mu\nu}$ that
\begin{equation}
\mathop{G^{\mu\nu}}\limits^{(0)}=2K^{\hphantom{\alpha}\mu}_{\alpha\hphantom{\nu}\rho}S^{\alpha\rho\nu}-2\mathop{\bigtriangledown_{\alpha}}\limits^{(0)}S^{\mu\alpha\nu}+\frac12 \mathbb T g^{\mu\nu}
\end{equation}
which shows in turn that, given the antisymmetry of the  contortion tensor, if the antisymmetric part of the  equations ($\left(S_{\mu\nu\alpha}-S_{\mu\nu\alpha}\right)\partial^{\alpha}{\mathbb T}=0$) is satisfied then so are the Bianchi identities.

\section{\label{sec:spherical} Spherically symmetric ans{\" a}tze}

In this Section we consider spherically symmetric solutions in vacuum. They were previously discussed in a number of papers, from different viewpoints. We want to show that two different existing approaches (with diagonal tetrad and non-zero spin connection and with non-diagonal ``good" tetrad) are actually fully equivalent to each other, and to explain in the next Section that the  corresponding equations are solvable precisely due to the Bianchi identities.

Spherically symmetric solutions are commonly searched for in spherical coordinates:
\begin{equation}
ds^2 = A(r)^2 dt^2 - B(r)^2 dr^2 - r^2[d\theta^2 + \sin^2(\theta)d\phi^2].
\label{sphAB}
\end{equation}
In terms of the tetrad, there are two different choices in the literature. 

First, this is the diagonal tetrad 
\begin{equation}
\label{dtetr}
E^{a}_{\mu} = \text{diag}(A(r),B(r),r,r\sin(\theta))
\end{equation}
which requires non-trivial spin connection \cite{Krssak:2018ywd}
\begin{equation}
\label{omeg}
\begin{split}
& \omega^{1}_{\ \theta 2} = - \omega^{2}_{\ \theta 1} = -1, \\ 
& \omega^{1}_{\ \phi 3} = -\omega^{3}_{\ \phi 1} = -\sin(\theta), \\
& \omega^{2}_{\ \phi 3}=-\omega^{3}_{\ \phi 2}=-\cos(\theta)
\end{split}
\end{equation}
for being consistent (other components vanish).

Second, it is the tetrad (for brevity, $s\equiv\sin$, $c\equiv\cos$) \cite{Daouda:2012nj,Bahamonde:2019zea}
\begin{equation}
\label{ndtetr}
E^{a}_{\mu} = \left(
\begin{array}{cccc}
A(r) & 0 & 0 & \\
0 & B(r) s(\theta)c(\phi) & r c(\theta) c(\phi) & -rs(\theta) s(\phi) \\
0 & B(r) s(\theta) s(\phi) & r c(\theta) s(\phi) & r s(\theta) c(\phi) \\
0 & B(r) c(\theta) & -rs(\theta) & 0
\end{array}
\right)
\end{equation}
which works with zero spin connection, and in the old non-covariant language it would be called a good tetrad which means that the antisymmetric part of equations is satisfied in the pure tetrad formalism.

Of course, one tetrad can be obtained from the other by a local Lorentz transformation. Indeed, the matrix
\begin{equation}
\label{lorentz}
\Lambda = \left(
\begin{array}{cccc}
1 & 0 & 0 & \\
0 &  s(\theta)c(\phi) &  c(\theta) c(\phi) & - s(\phi) \\
0 &  s(\theta) s(\phi) &  c(\theta) s(\phi) &  c(\phi) \\
0 &  c(\theta) & -s(\theta) & 0
\end{array}
\right)
\end{equation}
obviously transforms the tetrad (\ref{dtetr}) into  (\ref{ndtetr}). On the other hand, transforming back with $\Lambda^{-1}$ from the zero spin connection choice to the diagonal tetrad, we obtain the spin connection $\omega_{\mu}=-(\partial_{\mu}\Lambda^{-1})\Lambda=\Lambda^{-1}\partial_{\mu}\Lambda$. A back-of-an-envelope calculation shows that we get the spin connection (\ref{omeg}). In other words, those are just one and the same solution written in different Lorentzian frames.

One can substitute either the tetrad (\ref{ndtetr}) with zero spin connection or the ansatz (\ref{dtetr}, \ref{omeg}) into the equations of motion. We get the torsion scalar 
\begin{equation}
\label{torscal}
{\mathbb T} = -\dfrac{2  (B-1)(A-AB+2rA^{\prime}) }{r^2 A B^2}
\end{equation}
and the following relations
\begin{widetext}
\begin{eqnarray}
 {\mathfrak T}^t_t  & = & -\frac12 f - \dfrac{2}{r^2 A B^3}f_{T}\cdot\left(r(B-1)BA' +A(B^2-B+rB')\vphantom{\int} \right) \nonumber \\
 & & + \dfrac{8  (B-1)}{r^4 A^2 B^5} f_{TT}\cdot \left(\vphantom{\int}(B-1)\left(A^2(B^2-B-rB')-r^2B A'^2\vphantom{\int}\right) \right.\nonumber \\
& & \left. + rA\left(2rA'B' + B\left(A'(1-rB')-rA''\vphantom{A^A_A}\right)+B^2(-A'+rA'') \vphantom{\int}\right) \vphantom{\int}\right),\label{eqtime}  
\end{eqnarray}

\begin{eqnarray}
{\mathfrak T}^r_r & = & -\frac12 f - \dfrac{2}{r^2 A B^2}f_{T}\cdot \left(A(B-1)+rA'(B-2)\vphantom{\int} \right), \label{eqradius}\\
{\mathfrak T}^{\theta}_{\theta}  =  {\mathfrak T}^{\phi}_{\phi} & = & -\frac12 f + \dfrac{1}{r^2 A B^3} f_{T}\cdot\left(A\left(B-2B^2+B^3-rB'\vphantom{A^A_A}\right) +r\left(-2B^2A' - rA'B' + B(3A'+rA'')\vphantom{A^A_A}\right) \vphantom{\int}\right) \nonumber \\
& & - \dfrac{4(A-AB+rA')}{r^4 A^3 B^5}f_{TT}\cdot\left(-r^2 B A'^2(B-1) + A^2(B-1)(-B+B^2-rB')\vphantom{\int}\right.\nonumber \\
& & + \left.rA\left(2rA'B' +B\left(A'(1-rB')-rA''\vphantom{A^A_A}\right) + B^2(-A'+rA'')\vphantom{\int} \right) \vphantom{\int}\right).\label{eqangle}
\end{eqnarray}
\end{widetext}

Note that we have exploited the local Lorentz invariance of the covariant version of the theory. Independently, one can also use the diffeomorphism invariance. With another choice of the radial variable, such that the full spatial part of the metric is proportional to $B(r)$, all entries of the spatial part of the tetrad (\ref{ndtetr}) would be proportional to $B$, and then it is easy to see that they are nothing but components of Cartesian unit vectors in spherical coordinates. 

This is actually the geometric meaning of the Lorentz matrix (\ref{lorentz}): it describes the rotation between the spherical and Cartesian bases. It gives an idea that one could have diagonal tetrad with zero spin connection if using Cartesian coordinates. And indeed, we have checked that the diagonal tetrad for $ds^2=A^2(r)dt^2-B^2(r)\cdot (dx^2+dy^2+dz^2)$ with $r\equiv\sqrt{x^2+y^2+z^2}$ automatically satisfies the antisymmetric part of the  equations of motion, thus being a ``good'' tetrad in the old language.

\section{\label{sec:exist} Existence of solutions}

Equations (\ref{eqtime}, \ref{eqradius}, \ref{eqangle}) appeared previously in papers on spherically symmetric solutions in $f(\mathbb T)$ gravity. We can see that those ${\mathfrak T}^{\mu}_{\nu}=0$ are three equations for two variables. Obviously, for them being solvable without restricting the form of the function $f$ there must be dependence among them. Nevertheless, it is easy to convince oneself that there is no algebraic dependence. Even though previous authors mention that these equations do not restrict the functional form of $f$, it is not a priori obvious, especially given that very few exact solutions were known. Therefore, one of our main points in the current Letter is that there is no miracle behind the ability of people to solve this system, at least perturbatively, and the final answer is the (differential) Bianchi identities. 

Since the antisymmetric part of equations is automatically satisfied for our ans{\" a}tze, the Bianchi identities take the form
\begin{equation}
\label{BIsym}
\mathop{\bigtriangledown_{\mu}}\limits^{(0)}{\mathfrak T}^{\mu}_{\nu}=0.
\end{equation}
After a simple calculation of the Levi-Civita connection components for the metric \eqref{sphAB}, and taking into account that off-diagonal components of ${\mathfrak T}$ vanish, the $\nu=r$ component of \eqref{BIsym} reads:
\begin{equation}
\partial_r {\mathfrak T}^r_r+\left(\frac{A^{\prime}}{A}+\frac{2}{r}\right){\mathfrak T}^r_r-\frac{A^{\prime}}{A}{\mathfrak T}^t_t-\frac{1}{r}\left({\mathfrak T}^{\theta}_{\theta}+{\mathfrak T}^{\phi}_{\phi}\right)=0.
\end{equation}

One can check that this relation between the three equations \eqref{eqtime}, \eqref{eqradius} and \eqref{eqangle} holds indeed, reducing the number of independent equations for functions $A(r)$ and $B(r)$ to two. Finally, we need to solve for the radial equation \eqref{eqradius} which has less differential order, and any one combination of the two remaining equations.

In our opinion, a natural idea would be to combine equations (\ref{eqtime}) and (\ref{eqangle}), and to get rid of the $f_{TT}$ term. We obtain a simpler equation
\begin{equation}
\begin{split}
& f(\mathbb T)\cdot \left(A^2 B^2 r^2(B-1) + r^3 A A^{\prime} B^2\vphantom{\int}\right) \\
& + f_T(\mathbb T)\cdot \left( 4 A^2B(B-1) - 4A^2 B^2(B-1) + 4rA A^{\prime}\vphantom{\int}\right. \\ 
& \left. - 8r AA^{\prime}B + 4rAA^{\prime}B^2 
  + 4r^2 A^{\prime 2}(B-1)  + 4 r^2A A^{\prime} B^{\prime} \right. \\
  & \left. - 4 r^2 A A^{\prime\prime}(B-1) \vphantom{\int}\right) = 0. \label{compleq}
\end{split}
\end{equation}

We need to solve this equation (\ref{compleq}) together with the remaining radial equation (\ref{eqradius}) which can be written as
\begin{equation}
\label{eqrad}
  f(\mathbb T) + 4f_T(\mathbb T)\cdot\dfrac{-2 A^{\prime} r + AB + A^{\prime}Br-A}{A B^2 r^2}  = 0.
\end{equation}
Since the torsion scalar (\ref{torscal}) does not depend on derivatives of $B$, we can algebraically solve this equation for $B$ in terms of $A$. After that the equation \eqref{compleq} reduces to an ordinary differential equation for one single unknown function $A(r)$.

On the other hand, we can substitute $f(\mathbb T)$ found from equation (\ref{eqrad}) into the combined equation (\ref{compleq}), and see that as long as $f_T\neq 0$ we have the following relation between $A$ and $B$
\begin{equation}
-A^2(B+1)(B-1)^2 + r^2 A^{\prime 2} + r^2 A(A^{\prime} B^{\prime} - A^{\prime \prime}(B-1) )=0
\label{gencond}
\end{equation}
which is independent of the particular function $f(\mathbb T)$.

Let us recapitulate. The problem of finding spherically symmetric solutions in $f(\mathbb T)$ gravity is reduced to solving one algebraic equation (\ref{eqrad}) for $B$ in terms of $A$ and $A^{\prime}$, and then solving a second order ordinary differential equation for $A(r)$, which can be obtained by substituting the  solution for $B$ into one of remaining equations, for example equation (\ref{compleq}).

If a solution under consideration does not correspond to a (rather pathological, vanishing effective Planck mass) case of constant $\mathbb T$ with $f_T=0$ \footnote{Note that $f_T$ generically does not vanish even if $T=0$.}, then independently of the particular form of the function $f$ the functions $A$ and $B$ satisfy equation (\ref{gencond}). Note that it is not an independent condition; it follows from the equations of motion.

\subsection{Known exact solutions}

Unfortunately, even though for simple functions $f$ it is often possible to explicitly solve the algebraic equation for $B$ in terms of $A$, it appears very hard to find anything analytically for $A$ after that.

Let us consider a very simple function $f(\mathbb T)=\mathbb T^2$ which does not have well-defined $\mathbb T\to 0$ limit because of vanishing $f_T$. From equation \eqref{eqrad} we get four branches of solutions. 

Two of those branches are cases with  $\mathbb T=0$. One of them is simply $B=1$, and the other one is given by
\begin{equation}
B=\dfrac{A+2rA'}{A}.
\end{equation}
They can also be found by vanishing the torsion scalar \eqref{torscal}.
After that one can check that all other equations are satisfied automatically. In other words, any function $A(r)$ solves the equations given that $B$ is one of the above two choices. Another way to see the same is to have a look at the initial equations (\ref{coveom}) having in mind $f=f_T=\mathbb T=0$.

This degeneracy of solutions is the consequence of effectively switching off non-trivial dynamics of gravity for this model in the $\mathbb T\to 0$ limit. In healthier cases it (and any constant $\mathbb T$) would correspond to general relativity with a cosmological constant which previously already served well for finding solutions in $f(\mathbb T)$ gravity.

More interesting is to find solutions with non-constant torsion scalar. Those are two other branches of $\mathbb T^2$ gravity given by
\begin{equation}
B=\dfrac{-A-rA' \pm \sqrt{4A^2+8r A A' + r^2 A'^2} }{A}.
\end{equation}
Unfortunately, even for this simple case, the ODE for $A(r)$ appears very complicated.

We know one analytical solution from the work \cite{Bahamonde:2019jkf} of Bahamonde and Camci  . For $f(\mathbb T)=\mathbb T^n$ it reads $A(r)=r^{\alpha}$ with $\alpha=\frac{4n(n-1)(2n-3)}{4n^2-8n+5}$ and $B=\frac{(2n-1)(4n-5)}{4n^2-8n+5}$. It has non-constant torsion scalar which tends to zero when the radial variable is taken to infinity. Therefore, this solution is also pathological since it does not have a well-behaved asymptotic limit. Moreover, global properties require further scrutiny since a photon can reach $r=\infty$ in finite coordinate time $\propto \int\frac{dr}{r^{\alpha}}$ as long as $\alpha>1$.

\subsection{Asymptotic expansion method}

Since it is hard to find exact analytic solutions, it is interesting to study asymptotic limits far from the source. We assume Minkowski space at infinity
\begin{equation}
A(r) = 1 + \frac{a_1}{ r} + \frac{a_2}{ r^2} + \cdots,
\quad
B(r) = 1 + \frac{ b_1}{ r } + \frac{b_2} {r^2} + \cdots
\end{equation}
and want to find restrictions imposed by the general condition (\ref{gencond}) for any potentially healthy solution.

In the first non-trivial order ($\frac{1}{r^2}$) it gives
\begin{equation}
a_1^2 - a_1 b_1 - 2 b_1^2=0,
\end{equation}
which has two branches, $a_1=2b_1$ and $a_1=-b_1$ (no Birkhoff theorem). Having chosen one of them one can find precise values from equation (\ref{eqrad}), or any other remaining combination of equations of motion. Then at the next order of equation (\ref{gencond}) we will get unambiguously $b_2$ in terms of $a_2$ and the previous order solution, and so on.

Note that having obtained $a_1=-b_1$ is an important consistency check. The relation \eqref{gencond} which does not depend on the function $f$, and in particular must be valid for TEGR, must therefore admit the Schwarzschild solution, which we have also checked exactly.

\section{Conclusions}
\label{sec:concl}

We have shown that two different approaches to spherically symmetric solutions in $f(\mathbb T)$ gravity, preferring diagonal tetrad or zero spin connection, are fully equivalent if the local-Lorentz-covariant approach is employed. Moreover, switching to Cartesian coordinates instead of spherical ones can allow to meet both requirements. 

We have proven Bianchi identities for $f(\mathbb T)$ gravity and applied them to spherically symmetric solutions. It explains how the previous authors were able to solve three seemingly independent equations for two unknown functions. And we see that it is not a coincidence related to a choice of a simple function $f$ or to an artefact of a series expansion. It is a general property of equations of motion in the theory. 

Very importantly, our findings pave the way to better confronting this class of theories with observations. When a systematic study of spherically symmetric solutions is possible, one can explore phenomenological aspects of astrophysical compact objects, and in perspective to also apply it to top-hat collapse models, and to predict the halo mass function for studying the large scale structure in $f(\mathbb T)$ models.

\begin{acknowledgments}
The authors are grateful to Sebastian Bahamonde, Rafael Ferraro, Amr El-Zant and Waleed El Hanafy for useful discussions. M.J.G. was funded by FONDECYT-ANID postdoctoral grant 3190531. M.J.G. thanks the hospitality of colleagues at the Centre for Theoretical Physics of BUE, where part of this work was completed.
\end{acknowledgments}

\bibliography{apssamp}

\begin{thebibliography}{100}

\bibitem{Bull:2015stt}
P.~Bull {\it et.al}, Beyond $\Lambda$CDM: Problems, solutions, and the road ahead, Phys. Dark Univ. \textbf{12}, 56-99 (2016)

\bibitem{DelPopolo:2016emo}
A.~Del Popolo and M.~Le Delliou, Small scale problems of the $\Lambda$CDM model: a short review, Galaxies \textbf{5}, no.1, 17 (2017)

\bibitem{Bernal:2016gxb}
J.~L.~Bernal, L.~Verde and A.~G.~Riess, The trouble with $H_0$, JCAP \textbf{10}, 019 (2016)

\bibitem{BeltranJimenez:2019tjy}
J.~Beltr\'an-Jim\'enez, L.~Heisenberg and T.~S.~Koivisto, The Geometrical Trinity of Gravity, Universe \textbf{5}, no.7, 173 (2019)

\bibitem{Ferraro:2006jd}
R.~Ferraro and F.~Fiorini,  Modified teleparallel gravity: Inflation without inflaton, Phys. Rev. D \textbf{75}, 084031 (2007)

\bibitem{Bengochea:2008gz}
G.~R.~Bengochea and R.~Ferraro, Dark torsion as the cosmic speed-up, Phys. Rev. D \textbf{79}, 124019 (2009)

\bibitem{Ferraro:2008ey}
R.~Ferraro and F.~Fiorini, On Born-Infeld Gravity in Weitzenbock spacetime, Phys. Rev. D \textbf{78}, 124019 (2008)

\bibitem{Awad:2017sau}
A.~Awad and G.~Nashed, Generalized teleparallel cosmology and initial singularity crossing, JCAP \textbf{02}, 046 (2017)

\bibitem{Nunes:2018xbm}
R.~C.~Nunes, Structure formation in $f(T)$ gravity and a solution for $H_0$ tension, JCAP \textbf{05}, 052 (2018)

\bibitem{El-Zant:2018bsc}
A.~El-Zant, W.~El Hanafy and S.~Elgammal, $H_0$ Tension and the Phantom Regime: A Case Study in Terms of an Infrared $f(T)$ Gravity, Astrophys. J. \textbf{871}, no.2, 210 (2019)

\bibitem{Wang:2020zfv}
D.~Wang and D.~Mota, Can $f(T)$ gravity resolve the $H_0$ tension?, [arXiv:2003.10095 [astro-ph.CO]].

\bibitem{Golovnev:2018wbh}
A.~Golovnev and T.~Koivisto, Cosmological perturbations in modified teleparallel gravity models, JCAP \textbf{11}, 012 (2018)

\bibitem{Cai:2015emx}
Y.~F.~Cai, S.~Capozziello, M.~De Laurentis and E.~N.~Saridakis, f(T) teleparallel gravity and cosmology, Rept. Prog. Phys. \textbf{79}, no.10, 106901 (2016)

\bibitem{Ferraro:2011ks}
R.~Ferraro and F.~Fiorini, Spherically symmetric static spacetimes in vacuum f(T) gravity, Phys. Rev. D \textbf{84}, 083518 (2011)

\bibitem{Nashed:2014sea}
G.~G.~Nashed, Schwarzschild solution in extended teleparallel gravity, EPL \textbf{105}, no.1, 10001 (2014)

\bibitem{Nashed:2013bfa}
G.~G.~Nashed, Spherically symmetric charged-dS solution in $f(T)$ gravity theories, Phys. Rev. D \textbf{88}, 104034 (2013)

\bibitem{Nashed:uja}
G.~G.~Nashed, A special exact spherically symmetric solution in f(T) gravity theories, Gen. Rel. Grav. \textbf{45}, 1887-1899 (2013)

\bibitem{Bejarano:2014bca} 
  C.~Bejarano, R.~Ferraro and M.~J.~Guzm\'an, Kerr geometry in f(T) gravity, Eur.\ Phys.\ J.\ C {\bf 75}, 77 (2015)

\bibitem{Bejarano:2017akj} 
  C.~Bejarano, R.~Ferraro and M.~J.~Guzm\'an, McVittie solution in f(T) gravity, Eur.\ Phys.\ J.\ C {\bf 77}, no. 12, 825 (2017)

\bibitem{Paliathanasis:2014iva} 
A.~Paliathanasis, S.~Basilakos, E.~N.~Saridakis, S.~Capozziello, K.~Atazadeh, F.~Darabi and M.~Tsamparlis, New Schwarzschild-like solutions in f(T) gravity through Noether symmetries, Phys.\ Rev.\ D {\bf 89}, 104042 (2014)

\bibitem{Bahamonde:2019jkf}
S.~Bahamonde and U.~Camci, Exact Spherically Symmetric Solutions in Modified Teleparallel gravity, Symmetry \textbf{11}, no.12, 1462 (2019)

\bibitem{Boehmer:2019uxv}
C.~G.~B{\"o}hmer and F.~Fiorini, The regular black hole in four dimensional Born–Infeld gravity, Class. Quant. Grav. \textbf{36}, no.12, 12LT01 (2019)

\bibitem{Boehmer:2020hkn}
C.~G.~B{\"o}hmer and F.~Fiorini, BTZ gems inside regular Born-Infeld black holes, [arXiv:2005.11843 [hep-th]].

\bibitem{Bahamonde:2019zea}
S.~Bahamonde, K.~Flathmann and C.~Pfeifer, Photon sphere and perihelion shift in weak $f(T)$ gravity, Phys. Rev. D \textbf{100}, no.8, 084064 (2019)

\bibitem{Bahamonde:2020bbc}
S.~Bahamonde, J.~Levi Said and M.~Zubair, Solar System Tests in Modified Teleparallel Gravity, [arXiv:2006.06750 [gr-qc]].

\bibitem{Ortin2015} T. Ort\'in, Gravity and Strings, Cambridge Monographs on Mathematical Physics (Cambridge University Press, 2015)

\bibitem{Jimenez:2020dpn}
J.~Beltr\'an-Jim\'enez and A.~Delhom, Instabilities in metric-affine theories of gravity with higher order curvature terms, Eur. Phys. J. C \textbf{80} (2020) no.6, 585

\bibitem{Golovnev:2018red} 
  A.~Golovnev, Introduction to teleparallel gravities, Proceedings of the 9th Mathematical Physics Meeting: School and Conference on Modern Mathematical Physics, Institute of Physics, Belgrade 2018, pp. 219 - 236; arXiv:1801.06929 [gr-qc]
  
\bibitem{Bejarano:2019fii}
C.~Bejarano, R.~Ferraro, F.~Fiorini and M.~J.~Guzm\'an, Reflections on the covariance of modified teleparallel theories of gravity, Universe \textbf{5}, 158 (2019)

\bibitem{Krssak:2015oua}
M.~Kr\v{s}\v{s}\'ak and E.~N.~Saridakis, The covariant formulation of f(T) gravity, Class. Quant. Grav. \textbf{33}, no.11, 115009 (2016)

\bibitem{Golovnev:2017dox} 
  A.~Golovnev, T.~Koivisto and M.~Sandstad, On the covariance of teleparallel gravity theories, Class.\ Quant.\ Grav.\  {\bf 34}, no. 14, 145013 (2017)

\bibitem{Hohmann:2018rwf}
M.~Hohmann, L.~J{\"a}rv and U.~Ualikhanova, Covariant formulation of scalar-torsion gravity, Phys. Rev. D \textbf{97}, no.10, 104011 (2018).

\bibitem{Ferraro:2018tpu} 
R.~Ferraro and M.~J.~Guzm\'an, Hamiltonian formalism for f(T) gravity, Phys.\ Rev.\ D {\bf 97}, no. 10, 104028 (2018)

\bibitem{Li:2011rn}
M.~Li, R.~X.~Miao and Y.~G.~Miao, Degrees of freedom of $f(T)$ gravity, JHEP \textbf{07}, 108 (2011)

\bibitem{Pereira2012} R. Aldrovandi, J. G. Pereira, Teleparallel Gravity: an introduction, Springer, Dordrecht (2012).

\bibitem{Jimenez:2020ofm}
J.~Beltr\'an-Jim\'enez, A.~Golovnev, T.~Koivisto and H.~Veerm\"{a}e, Minkowski space in $f(T)$ gravity, [arXiv:2004.07536 [gr-qc]].

\bibitem{Ferraro:2020tqk}
R.~Ferraro and M.~J.~Guzm\'an, Pseudoinvariance and the extra degree of freedom in f(T) gravity, Phys. Rev. D \textbf{101}, no.8, 084017 (2020)

\bibitem{Golovnev:2019kcf}
A.~Golovnev and M.~J.~Guzman, Disformal transformations in modified teleparallel gravity, Symmetry \textbf{12}, no.1, 152 (2020)

\bibitem{Krssak:2018ywd}
M.~Krssak, R.~van den Hoogen, J.~Pereira, C.~B{\"o}hmer and A.~Coley, Teleparallel theories of gravity: illuminating a fully invariant approach, Class. Quant. Grav. \textbf{36}, no.18, 183001 (2019)

\bibitem{Daouda:2012nj} 
M.~H.~Daouda, M.~E.~Rodrigues and M.~J.~S.~Houndjo, Anisotropic fluid for a set of non-diagonal tetrads in f(T) gravity, Phys.\ Lett.\ B {\bf 715}, 241 (2012)

\end{thebibliography}

\end{document}